\makeatletter
\def\@sect#1#2#3#4#5#6[#7]#8{\ifnum #2>\c@secnumdepth
     \def\@svsec{}\else
     \refstepcounter{#1}\edef\@svsec{\csname the#1\endcsname.\hskip 1em }\fi
     \@tempskipa #5\relax
      \ifdim \@tempskipa>\z@
        \begingroup #6\relax
          \@hangfrom{\hskip #3\relax\@svsec}{\interlinepenalty \@M #8\par}
        \endgroup
       \csname #1mark\endcsname{#7}\addcontentsline
         {toc}{#1}{\ifnum #2>\c@secnumdepth \else
                      \protect\numberline{\csname the#1\endcsname}\fi
                    #7}\else
        \def\@svsechd{#6\hskip #3\@svsec #8\csname #1mark\endcsname
                      {#7}\addcontentsline
                           {toc}{#1}{\ifnum #2>\c@secnumdepth \else
                             \protect\numberline{\csname the#1\endcsname}\fi
                       #7}}\fi
     \@xsect{#5}}
\def\label#1{\@bsphack\if@filesw {\let\thepage\relax
   \xdef\@gtempa{\write\@auxout{\string
      \newlabel{#1}{{\thesection.\@currentlabel}{\thepage}}}}}\@gtempa
   \if@nobreak \ifvmode\nobreak\fi\fi\fi\@esphack}
\def\@eqnnum{(\thesection.\theequation)}
\documentstyle[12pt]{article}
\makeatletter
\def\section{\setcounter{equation}{0} \@startsection {section}{1}{\z@}{-3.5ex
   plus -1ex minus -.2ex}{2.3ex plus .2ex}{\Large\bf}}

\author{G. Jikia$^{(a)}$  and  A. Tkabladze$^{(b)}$ \\
[1ex]{${}^{(a)}$\it Institute for High Energy Physics},\\
     {\it 142284, Protvino, Moscow Region,}\\
     {\it Russian Federation}\\
[1ex]{${}^{(b)}$\it Kutaisi State University}\\
{\it 384000, Kutaisi, Georgia} }
\title{June 1993 \hfill IHEP 93--89 \\ \vspace*{2cm}
Photon-Photon Scattering at the Photon Linear Collider}
\date{}
\def\l{\lambda}

\def\g{\gamma}
\def\mw{m_W}

\def\Im{{\cal I \mskip-4.5mu \lower.1ex \hbox{\it m}}\,}
\def\Re{{\cal R \mskip-4mu \lower.1ex \hbox{\it e}}\,}
\def\IJMP #1 #2 #3 {{\it Int.\ J.\ Mod.\ Phys.}\ {\bf #1}\ (#2) #3}
\def\MPL #1 #2 #3 {{\it Mod.\ Phys.\ Lett.}\ {\bf #1}\ (#2) #3}
\def\NC #1 #2 #3 {{\it Nuovo Cim.}\ {\bf #1} (#2) #3}
\def\NP #1 #2 #3 {{\it Nucl.\ Phys.}\ {\bf #1}\ (#2) #3}
\def\PL #1 #2 #3 {{\it Phys.\ Lett.}\ {\bf #1}\ (#2) #3}
\def\PR #1 #2 #3 {{\it Phys.\ Rev.}\ {\bf #1}\ (#2) #3}
\def\PP #1 #2 #3 {{\it Phys.\ Rep.}\ {\bf #1}\ (#2) #3}
\def\PRL #1 #2 #3 {{\it Phys.\ Rev.\ Lett.}\ {\bf #1}\ (#2) #3}
\def\RMP #1 #2 #3 {{\it Rev.\ Mod.\ Phys.}\ {\bf #1}\ (#2) #3}
\begin{document}
\maketitle

\begin{abstract}
Photon-photon scattering at the Photon Linear Collider is considered. Explicit
formulas for helicity amplitudes due to $W$ boson loops are presented. It is
shown that photon-photon scattering should be easily observable at PLC and
separation of the $W$ loop contribution (which dominates at high energies)
will be possible at $e^+e^-$ c.m. energy of 500~GeV or higher.
\end{abstract}

\newpage

\section{Introduction}

Experiments on the observation of the processes involving photon-photon
collisions are extremely difficult to perform, because they are crossed-beam
experiments which require the highest intensities and the most sensitive
detection equipment. For this reason at the present time the photon-photon
cross sections are still of little interest to the experimental physicist. The
scattering of photons by a Coulomb field (Delbr\" uck scattering) is the only
observable case at the present time (see, {\it e.g.}, \cite{jarlskog}, review
articles \cite{delbruck} and references therein).

With the advent of new collider technique \cite{plc} the collision of high
energy, high intensity photon beams at the Photon Linear Collider (PLC),
obtained via Compton backscattering of laser beams off linac electron beams,
would provide novel opportunities for such processes. Based on the $e^+e^-$
linear collider PLC will have almost the same energy and luminosity, {\it i.e.}
c.m.  energy of 100-500 GeV and luminosity of the order of $10^{33}$ cm$^{-2}$
s$^{-1}$
\cite{plc}.

Explicit formulas for the fermion loop contribution to $\g\g\to\g\g$ are well
known \cite{cdtp}.  The additional charged $W$ boson loop contribution was
shown to be finite \cite{finite} but attracted little attention until recently,
when the $W$ loop contribution to the polarization tensor of photon-photon
scattering was calculated \cite{jiang,dong}. The resulting expressions of a
tedious calculation \cite{jiang} are too complicated to be implemented
numerically, and even the general properties like gauge invariance and Lorentz
covariance are only checked using low energy expansions. In addition, low
energy expressions obtained in \cite{dong} are not applicable above the $W$
threshold.

In Section~2 we will give explicit analytic results for the $W$ boson loop
contributions to the helicity amplitudes and their asymptotic expressions in
the high and low energy limits. Section~3 will contain numerical results. Cross
sections of the photon-photon scattering in monochromatic polarized $\g\g$
collisions  as well as cross sections calculated with account of photon
spectrum  will be given. It is shown that photon-photon scattering cross
section is large enough to be observable at the PLC and even separation of the
$W$ loop contribution will be possible at large enough energy. This fact is
of fundamental significance as both triple and quartic $W$ boson vertices
contribute. Finally, in Section~4, conclusions will be made.

\section{The $\g\g\to \g\g$ helicity amplitudes}

We use the reduction algorithm from Ref. \cite{vermaseren} to express the
$\g\g\to\g\g$ polarization tensor and helicity amplitudes in a canonical form
in terms of the set of basic scalar loop integrals.  All calculations were done
both in 't~Hooft-Feynman gauge and  non-linear $R_{\xi}$ gauge  for $\xi=1$
described in \cite{aazz}. The Feynman diagrams contributing to the process in
non-linear gauge are shown in Fig.~1. Symmetry and transversality properties of
the polarization tensor were explicitly checked. The algebraic calculations
were carried out using symbolic manipulation program FORM \cite{form}.

Extracting an overall factor $e^4/(4\pi)^2=\alpha^2$ in the definition of the
helicity amplitudes ${\cal M}_{\l_1\l_2\l_3\l_4}$, we find three independent
amplitudes:

\begin{eqnarray}
	{\cal M}_{++++}(s,t,u)	= &&
12-12 \Biggl(1+2 \frac{t}{s}\Biggr) B(t)
-12 \Biggl(1+2 \frac{u}{s}\Biggr) B(u)+24 \mw^2 \frac{t u}{s} D(t,u)
\nonumber \\
&&+\frac{4}{s} \Biggl(4 s-6 \mw^2-3 \frac{t u}{s}\Biggr)
\Biggl(2 t C(t)+2 u C(u)-t u D(t,u)\Biggr)
\nonumber \\
&&+8 (s-\mw^2) (s-3 \mw^2)
\Biggl(D(s,t)+D(s,u)+D(t,u)\Biggr)
; \\
	{\cal M}_{+++-}(s,t,u)	= &&
 -12+24 \mw^4 \Biggl(D(s,t)+D(s,u)+D(t,u)\Biggr)
\nonumber \\
&& +12 \mw^2 stu
\Biggl(\frac{1}{s^2}D(t,u)+\frac{1}{t^2}D(s,u)+\frac{1}{u^2}D(s,t)\Biggr)
\nonumber \\
&&
-24 \mw^2 \Biggl(\frac{1}{s}+\frac{1}{t}+\frac{1}{u}\Biggr)
\Biggl(s C(s)+t C(t)+u C(u)\Biggr)
; \\
	{\cal M}_{++--}(s,t,u) =&&
 -12+24 \mw^4 \Biggl(D(s,t)+D(s,u)+D(t,u)\Biggr).
\end{eqnarray}

The remaining five amplitudes can be  expressed
in terms of the independent ones by parity and Bose symmetry
\begin{eqnarray}
{\cal M}_{++-+}(s,t,u) &=&  {\cal M}_{+++-}(s,t,u),\nonumber \\
{\cal M}_{+-++}(s,t,u) &=&  {\cal M}_{+++-}(s,t,u),\nonumber \\
{\cal M}_{+-+-}(s,t,u) &=&  {\cal M}_{++++}(t,s,u),\nonumber \\
{\cal M}_{+--+}(s,t,u) &=&  {\cal M}_{++++}(u,t,s),\nonumber \\
{\cal M}_{+---}(s,t,u) &=&  {\cal M}_{+++-}(s,t,u),
\end{eqnarray}
here $\l_i$ denotes the polarization of particle $i$ and
\begin{equation}
s = (p_1+p_2)^2,\quad t = (p_2-p_3)^2, \quad u = (p_1-p_3)^2,\quad s+t+u=0.
\end{equation}
The scalar four-point functions are given by
\begin{eqnarray}
&&D(s,t) =  \\
&&\frac{1}{i\pi^2}\int
 \frac{d^4q}{\left(q^2-\mw^2\right)
\left(\left(q+p_1\right)^2-\mw^2\right)
\left(\left(q+p_1+p_2\right)^2-\mw^2\right)
\left(\left(q+p_4\right)^2-\mw^2\right)}, \cdots\nonumber
\end{eqnarray}
and two- and three-point functions $B$ and $C$ are defined by analogous
expressions. The expressions for $B$, $C$ and $D$ functions in terms of (di-)
logarithms are well known \cite{cdtp}.

The unexpected fact is that helicity amplitudes ${\cal M}_{+++-}$, ${\cal
M}_{++--}$ etc., describing photon-photon scattering when algebraic sum of the
photon helicities is not conserved, are just $-3/2$ times fermion loop
contribution to the corresponding amplitudes for fermion mass equal to $\mw$.
The helicity conserving amplitudes ${\cal M}_{++++}$ etc. are different for $W$
and fermion loop contributions. Moreover, we found that $W$ loop contribution
to helicity non-conserving amplitudes were just 3 times the charged scalar loop
contribution. So, if we consider photons interacting with ``supersymmetric''
set of charged scalar, fermion and vector fields degenerate in mass ({\it
i.e.}, if the number of bosonic degrees of freedom of charged fields is
equal to the number of fermionic ones), helicity non-conserving amplitudes will
add exactly to zero and only helicity conserving amplitudes will survive!
It seems, this should be explained by the embedding of the Yang-Mills
theory in a supersymmetric theory.

In the low energy limit $s\ll\mw^2$ the leading terms of the expansion of
helicity amplitudes are
\begin{equation}
{\cal M}_{++++} = \frac{14}{5}\frac{s^2}{\mw^4},\quad
{\cal M}_{++--} = \frac{1}{10}\frac{s^2+t^2+u^2}{\mw^4},\quad
{\cal M}_{+++-} = 0, \cdots,
\end{equation}
and  simple closed expressions are easily obtained for the differential cross
section
\begin{equation}
\frac{d\sigma^W}{d\cos\theta}=\frac{393\alpha^4}{25600\pi}\frac{s^3}{\mw^8}
\left(1+\cos^2\theta\right)^2
\end{equation}
and total cross section
\begin{equation}
\sigma^W_{tot}=\frac{2751\alpha^4}{16000\pi}\frac{s^3}{\mw^8}
\approx 0.055\alpha^4\frac{s^3}{\mw^8},
\end{equation}
in precise agreement with the low energy expansion result of \cite{dong}.

The ratio of the electron and $W$ loop induced cross sections at
$\sqrt{s}\approx 3 m_e$, where $\sigma^e$ has a maximum of 1.6~$\mu$b
\cite{cdtp}, is extremely small $\sigma^W/\sigma^e\approx 115
(m_e/\mw)^8\approx 10^{-40}$. That is one of the reasons why $W$ loop
contribution has been ignored so far.

In the high energy limit with $u$ held finite and large, $s\gg -u\gg  \mw^2$,
which corresponds to photon scattering at small angle, $\theta\sim
(-u/s)^{1/2}$, only two helicity conserving amplitudes survive, which are
imaginary and proportional to $s$
\begin{equation}
{\cal M}_{++++}={\cal M}_{+-+-}=-16\pi i\frac{s}{u}\log(-u/\mw^2),
\label{asymptotics}
\end{equation}
while leading fermion loop contributions to helicity amplitudes
grow  only as logarithm squared \cite{akhiezer,cdtp}
\begin{equation}
{\cal M}_{++++}^f={\cal M}_{+-+-}^f=-4 \log^2(-s/u).
\end{equation}
It means that at high energies the $W$ loop contribution will not only be
non-negligible, but will even dominate in the total cross section of
photon-photon scattering. The fact that $W$ loop contribution gives rising
amplitudes and non-decreasing total cross section in the high energy limit is a
consequence of the $u$, $t$-channel vector particle exchanges \cite{gorshkov}.
Such behaviour can be qualitatively understood by considering amplitudes for
forward (backward) scattering that can easily be calculated via optical
theorem
\begin{equation}
\alpha^2 \Im {\cal M}^W_{++++(+-+-)}(u=0)=
s\sigma_{\gamma\gamma\to W^+W^-}^{++(+-)}=8\pi\alpha^2 s/\mw^2.
\end{equation}

\section{Cross sections}

Fig.~2 presents polarized cross sections for photon-photon scattering in
monochromatic $\g\g$ collisions summed over final photon helicities. We
consider here the extreme cases of $\l_1\l_2=\pm 1$, {\it i.e.} full circular
polarization for the incoming photons.  The cross section is given by
\begin{equation}
\frac{d\sigma_{\l_1,\l_2}(s)}{d\cos\theta} =\sum_{\l_3,\l_4}
\frac{\alpha^4}{32\pi s} \biggl|{\cal M}_{\l_1\l_2\l_3\l_4}\biggr|^2,
\end{equation}
where the integration over $\cos\theta$ should be done from 0 to 1. The
parameters  $\alpha = 1/128$, $m_W=80.22$ GeV and $m_t=120$ GeV have been used
throughout the paper. We present total cross sections as well as separate
contributions coming from $W$-boson loop and fermion loop. To get an idea of an
observable cross section, we restrict the photon scattering angle,
$|\cos\theta|<\cos 30^o$. As expected, below $W$ threshold the fermion loop
contribution is dominating, while $W$ loop dominates at photon-photon
collision energies above $200\div 250$~GeV.

In Fig.~3 we show the total cross sections of the photon-photon scattering
calculated with account of the photon spectrum \cite{plc} as a function of
the $e^+e^-$ c.m. energy for unpolarized initial electron and laser beams. The
large cross section originating from fermion loop contribution is due to a low
energy tail of the photon distribution function. {\it E.g.}, for $e^+e^-$ c.m.
energy below 200~GeV the  cross section is of the order of 100~fb, which
corresponds to about a thousand of events for the integrated luminosity of
10~fb$^{-1}$. To be prudent we also calculated  the background from resolved
photon contribution $q\bar q\to\gamma\gamma$. The invariant mass of two photons
was taken to be higher than 20~GeV.  We used the parametrization of photonic
parton distributions \cite{grv}. The resolved photon contribution is sizable,
but it also comes from low energy photon-photon pair production.

Fig.~4 shows that if one cuts off photon pairs with low invariant
mass it will be possible to separate the $W$ loop contribution. For
$M_{\g\g}>250$~GeV at $e^+e^-$ c.m. energy of 500~GeV the $W$ contribution to
cross section is dominating, $\sigma_W\approx 5$~fb (which corresponds to 50
events for $\int d{\cal L}=10$~fb$^{-1}$), while the fermion contribution is
negligible, $\sigma_f\approx 0.2$~fb. And the resolved photon contribution is
completely negligible for $M_{\g\g}>100$~GeV.

Future detectors possibilities to detect two photon production reactions were
carefully studied for rare two-photon decay of the Higgs particle \cite{hgg}.
The achievable precision of the photon angular measurements and two photon
invariant mass resolutions were shown to be 1~mrad and 1-3~GeV, respectively
\cite{hgg}. In our case much simpler electromagnetic calorimeter would be
enough to study photon-photon scattering reaction.  In practice one of the
major sources of background is the fragmentation of jet into a leading neutral
meson ($\pi^0$, $\eta^0$, or $K_L^0$). Fortunately, we can use the value of
current rejection factor of $5\times 10^{-4}$ obtained by CDF in their single
photon analysis \cite{cdf}, which is defined as the ratio of background in
their inclusive single photon cross section to the total inclusive jet cross
section. The jet cross section in $\gamma\gamma$ collisions for jet-jet
invariant mass above 100~GeV in the $p_t$ region above 30~GeV can be estimated
to be below 100~pb. Multiplying this cross section by the rejection factor
squared we obtain that this background is really negligible.

For the case of polarized electron and laser photon beams the low energy photon
tail can be significantly suppressed by varying the value of the conversion
distance $z$ (distance from the conversion point -- where the laser pulse
intersects the electron beam -- to the interaction point) \cite{plc}. Plotted
in Fig.~5 are the total cross sections as a function of the $e^+e^-$ c.m.
energy for polarized electron and laser beams and a conversion distance
$z=2.5$~cm. The $W$-loop contribution clearly dominates over fermion loop
contribution without any $\gamma\gamma$ invariant mass cut, while the value of
the cross section due to $W$-loop is only 20\% smaller than that for zero
conversion distance (Fig.~3). The resolved photon contribution is about 2~fb at
$e^+e^-$ c.m. energy of 500~GeV and can be further reduced by photon-photon
invariant mass cut (Fig.~4).

\section{Conclusions}

In conclusion, photon-photon scattering should easily be observable at the
PLC. Fermion loop contribution dominates below the $W$ threshold, but at
photon-photon collision energies above $200\div 250$~GeV the $W$ loop
contribution is dominating. For $e^+e^-$ c.m. energy of 500~GeV or higher it
will be possible to separate the $W$ contribution. The observation of the
photon-photon scattering due to the $W$ loop will have  fundamental
significance, because this process is a pure one-loop effect of the Standard
Model as a renormalizable nonabelian gauge theory. In principle, this reaction
can be used to probe the anomalous triple and quartic $W$ boson vertices.

\vspace*{1cm}
We are grateful to S.S.~Gershtein for valuable discussions and support. This
work was supported, in part, by the International Science Foundation Grant.

\newpage
\section*{Figure captions}
\parindent=0pt
\parskip=\baselineskip

Fig.~1. Feynman graphs contributing to the process $\g\g\to \g\g$.
Notations are the following: photon -- wavy line, $W$ boson -- zigzag line,
charged NG scalar -- solid line, FP ghost -- dashed line.

Fig.~2. Total cross section of photon-photon scattering in monochromatic
photon-photon collisions versus $\g\g$ c.m. energy for different helicities of
the incoming photons. Total cross section (solid line) as well as $W$ boson
loop contribution (dashed line) and fermion loop contribution (dotted line) are
shown.

Fig.~3. Cross section of photon-photon scattering in $\g\g$ collisions versus
c.m. energy of the $e^+e^-$ collisions computed taking into account photon
spectrum of the backscattered laser beams. Dash-dotted line shows the resolved
photon contribution.

Fig.~4. The invariant mass, $M_{\g\g}>M_{cut}$, distribution versus $M_{cut}$
for $\g\g\to \g\g$ at $\sqrt{s_{e^+e^-}}=500$~GeV for polarized electron and
laser beams.  Notations are the same as in Fig.~3.

Fig.~5.  Cross section of photon-photon scattering in $\g\g$ collisions versus
c.m. energy of the $e^+e^-$ collisions for polarized electron and laser beams.
The electron beam spotsize is 100~nm and the conversion distance is $z=2.5$~cm.
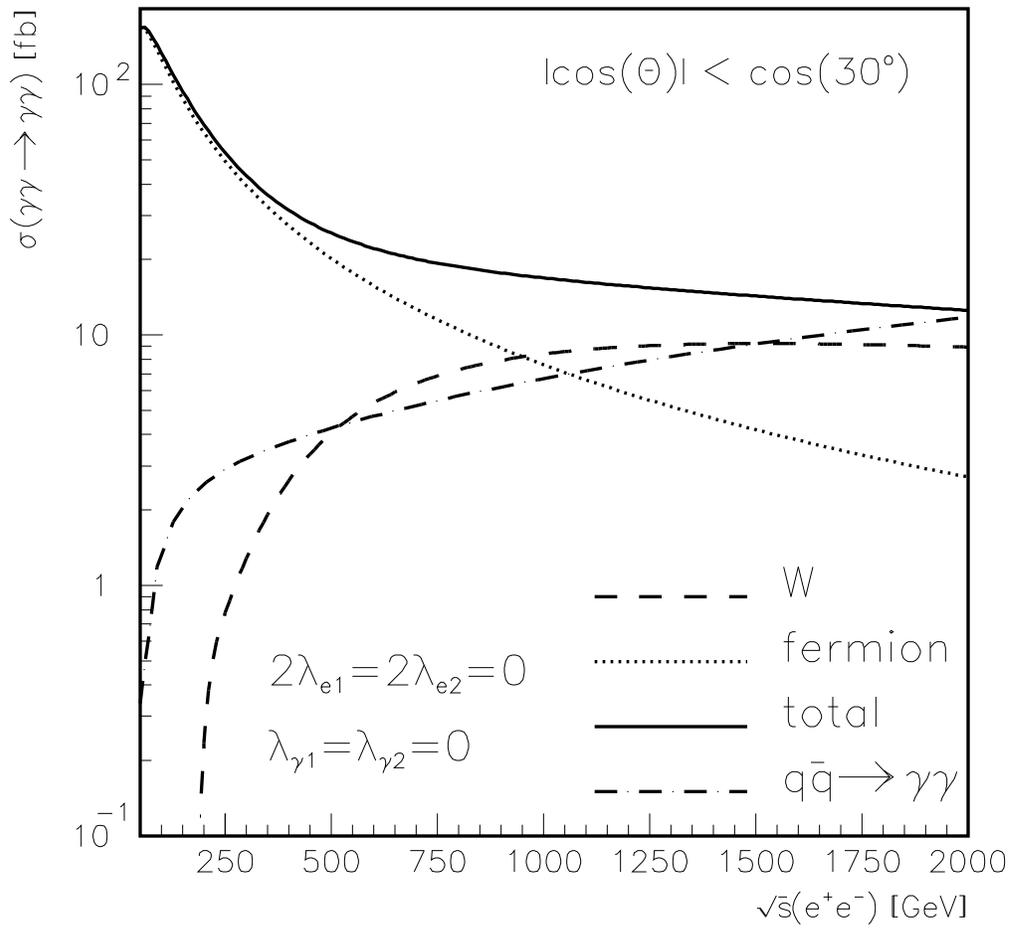
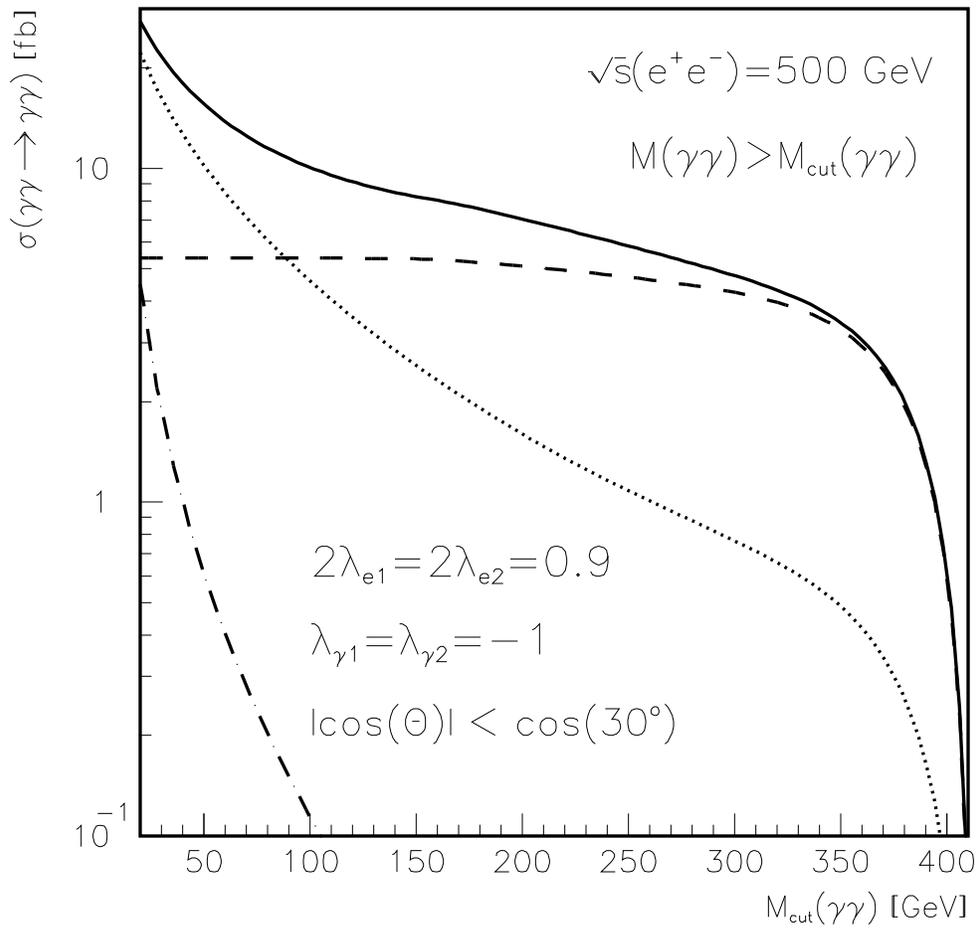
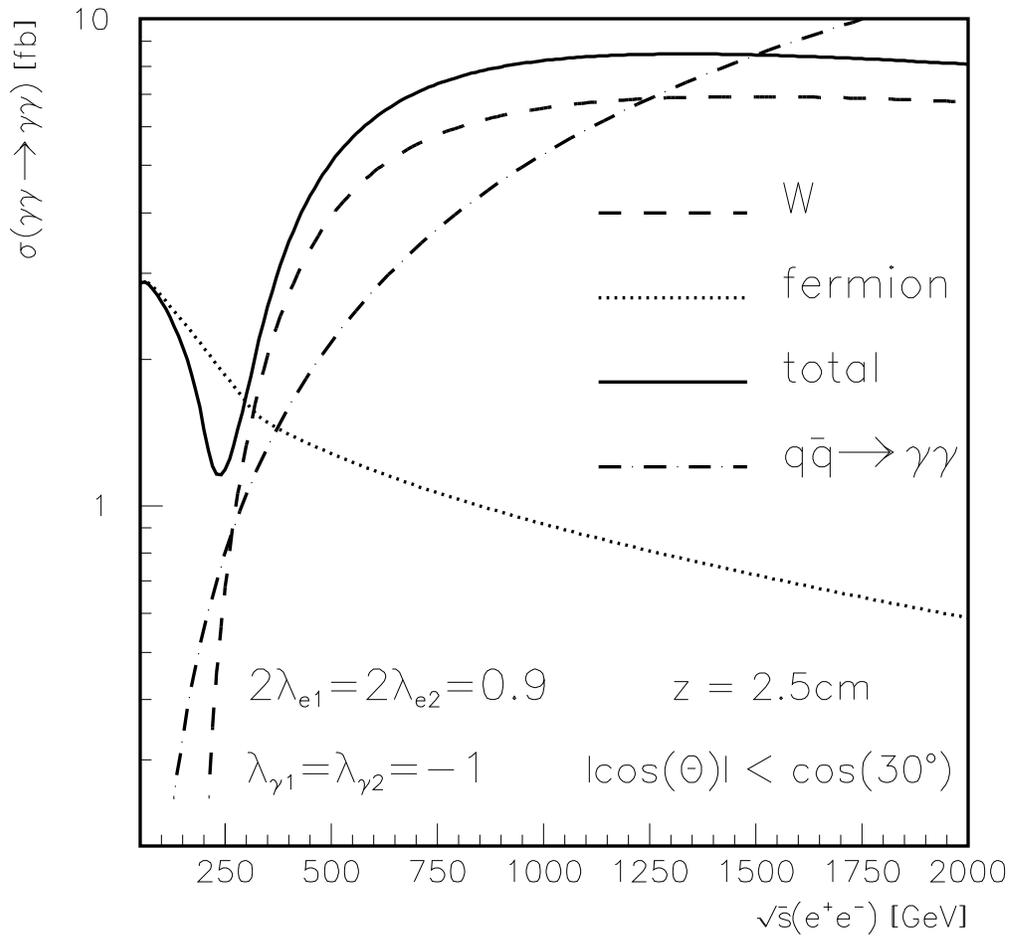
\end{document}